\documentclass[fleqn,3p,onecolumn,12pt]{elsarticle_arxiv0909.0160v4}

\newcommand{\versionnumber}{v4}
\journal{Physics Letters B 682 (2009) 316}
\renewcommand{\today}{arXiv:0909.0160 [hep-th] (\versionnumber)}

\usepackage{graphicx}
\usepackage{bm}
\usepackage{mathrsfs}
\usepackage[latin1]{inputenc}
\usepackage{stmaryrd}
\usepackage{array}
\usepackage{psfrag}
\usepackage{dsfont}
\usepackage{epsfig}
\usepackage{titletoc}
\usepackage{float}   
\usepackage{wrapfig} 
\usepackage{amsmath}
\usepackage[latin1]{inputenc}
\usepackage{amsmath}\usepackage{amssymb,stmaryrd}
\usepackage{mathrsfs}
\usepackage{array}
\usepackage{graphicx}
\usepackage{psfrag}
\usepackage{dsfont}
\usepackage{epsfig}
\usepackage{cancel}

\def\be {\begin{equation}}
\def\ee {\end{equation}}
\def\ba {\begin{eqnarray}}
\def\ea {\end{eqnarray}}

\newcommand{\bra}[1] {\left(#1\right)}
\newcommand{\vierbein}[2]{e_{#1}^{\;\;#2}}
\newcommand{\inversevb}[2]{e^{#1}_{\;\;#2}}

\begin{document}

\begin{frontmatter}

\title{Lorentz violation and black-hole thermodynamics:\\
       Compton scattering process}

\author{E. Kant} \ead{kant@particle.uni-karlsruhe.de}
\author{F.R. Klinkhamer\corref{cor1}}
\cortext[cor1]{Corresponding Author}\ead{frans.klinkhamer@kit.edu}
\author{M. Schreck} \ead{schreck@particle.uni-karlsruhe.de}
\address{Institute for Theoretical Physics, University of Karlsruhe,\\
         Karlsruhe Institute of Technology, 76128 Karlsruhe, Germany}

\begin{abstract}
A Lorentz-noninvariant modification of quantum electrodynamics (QED) is considered,
which has  photons described by the nonbirefringent sector of modified Maxwell
theory and electrons described by the standard Dirac theory.
These photons and electrons are taken to propagate and interact
in a Schwarzschild spacetime background.
For appropriate Lorentz-violating parameters,
the photons have an effective horizon lying outside the
Schwarzschild horizon.
A particular type of Compton scattering event, taking place
between these two horizons (in the photonic ergoregion)
and ultimately decreasing the mass of the black hole,
is found to have a nonzero probability.
These events perhaps allow for a violation of the generalized second law
of thermodynamics in the Lorentz-noninvariant theory considered.
\end{abstract}
\end{frontmatter}

\section{Introduction}

Lorentz-violating theories coupled to gravity can have interesting
black-hole solutions. Particles that obey Lorentz-violating
dispersion relations
 may perceive an effective horizon different from         
the event horizon for standard Lorentz-invariant
matter~\cite{DubovskySibiryakov2006,Eling-etal2007,Betschart-etal2009}. It
has been argued~\cite{DubovskySibiryakov2006,Eling-etal2007} that such
multiple-horizon structures allow for the construction of a perpetuum mobile
of the second kind (involving heat transfer from a cold body to a hot body,
without other change).

This Letter considers modified Maxwell theory~\cite{KosteleckyMewes2002}
as a concrete realization
of a Lorentz-violating theory. With an appropriate choice for the
Lorentz-violating parameters, the nonstandard photons have an effective
horizon lying outside the Schwarzschild event horizon for standard matter.
Of interest, now, are Compton scattering events  
$\gamma e^{-} \to \gamma e^{-}$, which take place between these two horizons,
that is, in the accessible part of the photonic ergosphere region.
After the collision, the photon may carry negative Killing energy
as it propagates inside the photonic ergosphere, so that the
final electron carries away more Killing energy than the sum of
the Killing energies of the ingoing particles.
As shown in Sec.~IV--B  of Ref.~\cite{Eling-etal2007}, such a scattering
event ultimately \emph{reduces} the black-hole mass.  In the following,
it will be demonstrated that this particular Compton scattering
event is kinematically allowed and has a nonvanishing probability to occur.

The purpose of this Letter is to give a concrete example
of a Compton scattering event that can be used to reduce the
black-hole mass. This requires a detailed discussion of the theory
in Sec.~\ref{sec:Setup}, which can, however, be skipped in a first
reading. The main result is presented in Sec.~\ref{sec:Compton}
and discussed in Sec.~\ref{sec:Discussion},
both of which sections are reasonably self-contained.

\setcounter{equation}{0}
\section{Setup}
\label{sec:Setup}
\subsection{Units and conventions}

Natural units are used with $c=G_\text{N}=\hbar =1$.
Spacetime indices are denoted  by Greek letters and correspond to
$t,r,\theta, \phi$ for standard spherical
Schwarzschild coordinates or to $\tau,R,\theta, \phi$ for Lema\^{i}tre coordinates.
Local Lorentz indices are denoted by Latin letters and run from $0$ to $3$.
The flat-spacetime Minkowski metric is $\eta_{a b}$
and the curved-spacetime Einstein metric $g_{\mu\nu}$,   
both with signature $\left(+,-,-,-\right)$.
The determinant of the metric is denoted  by $g\equiv \text{det}\,g_{\mu\nu}$.
The vierbeins are introduced in the standard way by writing
$g_{\mu\nu}=\vierbein{\mu}{a}\vierbein{\nu}{b}\,\eta_{ab}$
and obey the relations $\inversevb{\mu}{a}\vierbein{\mu}{b} = \delta_a^{~b}$
and $\inversevb{\mu}{a}\vierbein{\nu}{a} = \delta_{~\nu}^{\mu}$.

\subsection{Modified QED in curved spacetime}

Modified Maxwell theory is an Abelian $U(1)$ gauge theory
with a Lagrange density that consists of the standard Maxwell term and an
additional Lorentz-violating bilinear
term~\cite{KosteleckyMewes2002,BaileyKostelecky2004,
KlinkhamerRisse2008,KlinkhamerSchreck2008}.
The vierbein formalism is particularly well-suited for
describing Lorentz-violating theories in curved spacetime,
since it allows to distinguish between  local
Lorentz and general coordinate transformations~\cite{Kostelecky2004}
and to set the torsion identically to zero.

A minimal coupling procedure
then yields the following Lagrange density for the photonic part of the action:
\begin{subequations}\label{eq:LagMMgrav}
\begin{eqnarray}
\mathcal{L}_\text{modM}
&=&
  -\frac{1}{4}\,  g^{\mu\rho}g^{\nu\sigma}\, F_{\mu\nu}F_{\rho\sigma}
  -\frac{1}{4}\, \kappa^{\mu\nu\rho\sigma}\, F_{\mu\nu}F_{\rho\sigma} \ ,
\label{eq:LagMMgrav-L}
\\[2mm]
 \kappa^{\mu\nu\rho\sigma}
 &\equiv&
  \kappa^{abcd}\,e^\mu_{\;\;a}\,
  e^\nu_{\;\;b}\,e^\rho_{\;\;c}\,e^\sigma_{\;\;d}\,,
\label{eq:LagMMgrav-kappa-tensor}
\end{eqnarray}
\end{subequations}
in terms of the standard Maxwell field strength tensor
$F_{\mu\nu}\equiv \partial_\mu A_\nu -\partial_\nu A_\mu$.
The  ``tensor'' $\kappa^{abcd}$ has the same symmetries as
the Riemann curvature tensor, as well as a double-trace condition.
The numbers $\kappa^{abcd}(x)$ are considered to be fixed parameters,
with no field equations of their own.

In the following, we explicitly choose this background tensor field to be
of the form~\cite{BaileyKostelecky2004}
\begin{equation}\label{eq:ansatznonbire}
\kappa^{abcd}(x)=\frac{1}{2} \left(\eta^{ac}\,
\widetilde{\kappa}^{bd}(x)-\eta^{ad}\, \widetilde{\kappa}^{bc}(x)
     +\eta^{bd}\, \widetilde{\kappa}^{ac}(x)-\eta^{bc}\, \widetilde{\kappa}^{ad}(x)\right),
\end{equation}
in terms of a symmetric and traceless background field
$\widetilde{\kappa}^{ab}(x)$.
Physically, \eqref{eq:ansatznonbire} implies the restriction to the
nonbirefringent sector of modified Maxwell theory.
Moreover, we employ the following decomposition of $\widetilde\kappa^{ab}(x)$:
\begin{equation}\label{eq:ansatzxi}
\widetilde{\kappa}^{ab}(x)=\kappa\Big(\xi^a(x)\,\xi^b(x)-\eta^{ab}/4 \Big),
\end{equation}
relative to a normalized parameter four-vector $\xi^a$ with
$\xi_a \xi^a=1$.
For our purpose, we will choose the parameter $\kappa$
in \eqref{eq:ansatzxi} to be spacetime independent.

The breaking of Lorentz invariance in the electromagnetic theory
\eqref{eq:LagMMgrav}
is indicated by the fact that  
the flat-spacetime theory allows for maximal photon velocities
different from $c=1$ (operationally defined by the maximum attainable
velocity of standard Lorentz-invariant particles to be discussed shortly).
See, e.g., Refs.~\cite{KosteleckyMewes2002,BaileyKostelecky2004,
KlinkhamerRisse2008,KlinkhamerSchreck2008}
for further details of the simplest version of modified Maxwell theory
with constant $\kappa^{abcd}$ over Minkowski spacetime
and physical bounds on its 19 parameters.

The charged particles (electrons) are described by the
standard Dirac Lagrangian over curved spacetime~\cite{Birrell}
and gravity itself by the standard Einstein--Hilbert Lagrangian~\cite{Wald1984}.  
All in all, this particular modification of
quantum electrodynamics (QED) has action
\begin{subequations}\label{eq:fullaction-LagEH-LagD}
\begin{eqnarray}
S &=& \int_{\mathbb{R}^4} d^4x\;\sqrt{-g}\;
\big(\mathcal{L}_\text{EH}+\mathcal{L}_\text{D}+\mathcal{L}_\text{modM}\big),
\label{eq:fullaction}\\[1mm]
\mathcal{L}_\text{EH} &=& R/(16\pi)\,,
\label{eq:LagEH} \\[1mm]
\mathcal{L}_\text{D} &=& \overline{\psi} \left( \frac{1}{2}\, \gamma^a
e^\mu_{\;\;a}
\;  
\text{i} \overset{\leftrightarrow}{\nabla}_\mu-m\right) \psi\,,
\label{eq:LagD}
\end{eqnarray}
\end{subequations}
with  Ricci curvature scalar $R$ from the metric $g_{\mu\nu}$,
the usual Dirac matrices $\gamma^a$,
and the gauge- and Lorentz-covariant derivative of a spinor~\cite{Birrell},
\begin{subequations}
\begin{equation}
\nabla_\mu\psi \equiv \partial_\mu\,  \psi +\Gamma_\mu \, \psi-e A_\mu\,  \psi\;,
\end{equation}
with spin connection
%
%
\begin{eqnarray}  
\Gamma_\mu &=& \frac{1}{2}\, \Sigma^{ab}\, e_a^{\;\;\nu}\partial_\mu (e_{b\;
\nu}) \,,\quad \Sigma_{ab} \equiv \frac{1}{4}\left(\gamma_a\gamma_b
-\gamma_b\gamma_a\right)\,.
\end{eqnarray}
\end{subequations}

\subsection{Effective background for the photons}

As demonstrated in Sec.~3 of Ref.~\cite{Betschart-etal2009}, photons described by the
Lagrange density \eqref{eq:LagMMgrav}  with the Lorentz-violating parameters
\eqref{eq:ansatznonbire}--\eqref{eq:ansatzxi} propagate on null-geodesics of
an effective metric.
This effective metric is given by:
\begin{eqnarray}
\widetilde{g}_{\mu\nu}(x)&=&
g_{\mu\nu}(x)-\frac{\kappa}{1+\kappa/2}\;
\xi_{\mu}(x)\xi_{\nu}(x)\,,
\label{eq:effectivemetric}
\end{eqnarray}
with an inverse following from
$\widetilde{g}^{\mu\nu}\widetilde{g}_{\nu\rho}=\delta^{\mu}_{\;\;\rho}$.
All lowering or raising of
indices is, however, understood to be performed by contraction with
the original background metric $g_{\mu\nu}$ or its inverse $g^{\mu\nu}$,
unless stated otherwise.

In order to avoid obvious difficulties with causality, we
restrict our considerations to a subset of
theories without space-like  
photon trajectories (with respect to the
original metric). This is ensured by the choice $0\leq\kappa< 2$.

\subsection{Schwarz\-schild spacetime metric}

In the following, we consider a standard Schwarzschild geometry as given
by the following line element:
\begin{subequations}\label{eq:SS-lineelementstandard}
\begin{eqnarray}
ds^2 &=&\left(1-2 M/r\right) dt^2-\left(1-2 M/r\right)^{-1}
dr^2 -r^2 d\Omega^2\,,\\
d\Omega^2 &\equiv& d\theta^2+\sin^2\theta\, d\phi^2\,.
\end{eqnarray}
\end{subequations}
It will be convenient to work with Lema\^{i}tre coordinates,
\begin{equation}\label{eq:SS-lineelementlemaitre}  
ds^2 = d\tau^2-\left(\frac{3(R-\tau)}{4 M}\right)^{-2/3}dR^2
-\Big(3/2\, (R-\tau)\Big)^{4/3}(2 M)^{2/3}\,d\Omega^2\,,
\end{equation}
as Lema\^{i}tre coordinates describe the standard Schwarzschild solution in
coordinates which are nonsingular at the horizon (corresponding to the
reference frame of a free-falling observer).

The transformation to standard Schwarzschild coordinates reads
\begin{subequations}
\begin{equation}
d \tau=dt+\frac{\sqrt{2M/r}}{1-2M/r}\;dr\;,
\end{equation}
\begin{equation}
dR=dt+\frac{1}{\left(1-2M/r\right)\sqrt{2M/r}} \;dr\;,
\end{equation}
\end{subequations}
and the horizon is described by $\left(R-\tau\right)=(4/3)\, M$.
 A suitable choice of the vierbein $e_\mu^{\;a}$ is given by
%
%
\begin{eqnarray}\label{eq:vierbein}  
e_{\tau}^{\;0} &=& 1\,,\quad
e_R^{\;1}      = \sqrt{\left|g_{RR}\right|}\,,\quad
e_\theta^{\;2} = \sqrt{\left|g_{\theta\theta}\right|}\,,\quad
e_{\phi}^{\;3} = \sqrt{\left|g_{\phi\phi}\right|}\,,
\end{eqnarray}
with all other components vanishing.

\subsection{Effective Schwarz\-schild metric for the photons}

For the vector field $\xi^\mu(x)=e^\mu_{\;\;a}(x)\,\xi^a(x)$ entering the
nonstandard part of the
photonic action \eqref{eq:LagMMgrav}--\eqref{eq:ansatzxi}   
and the effective Lorentz-violating parameter, we take
\begin{subequations}\label{eq:xiexplicit-epsilon}
\begin{eqnarray}
\xi^\mu(x) &=& \left(1,0,0,0\right)\,,
\label{eq:xiexplicit}\\
\epsilon &\equiv& \frac{\kappa}{1-\kappa/2}\;,
\label{eq:epsilon}
\end{eqnarray}
\end{subequations}
where the first expression (in Lema\^{i}tre coordinates)
makes clear that the photonic Lorentz violation  
is isotropic
and the last expression introduces a convenient Lorentz-violating parameter
for the theory considered. The particular parameter choices
\eqref{eq:xiexplicit-epsilon} correspond to Case 1 in Ref.~\cite{Betschart-etal2009}.
Asymptotically ($R\to\infty$ for fixed $\tau$),
the parameter $\kappa$ corresponds to
$2\,\widetilde{\kappa}_\text{tr}$, in terms of the parameter
$\widetilde{\kappa}_\text{tr}$ introduced by
Ref.~\cite{KosteleckyMewes2002} and bounded in Ref.~\cite{KlinkhamerSchreck2008}.

As shown in Sec.~4.1 of Ref.~\cite{Betschart-etal2009}, the effective background for
the photons \eqref{eq:effectivemetric}  is again a  Schwarz\-schild
background,
%
%
\begin{equation}\label{eq:SS-lineelementlemaitremodmass} 
d\widetilde{s}^2=d\widetilde{\tau}^2-\left(\frac{3(\widetilde{R}-\widetilde{\tau})}
{4\widetilde{M}}\right)^{-2/3}d\widetilde{R}^2
-\Big(3/2\, (\widetilde{R}-\widetilde{\tau})\Big)^{4/3}(2 \widetilde{M})^{2/3}\,d\Omega^2\,,
\end{equation}
with a rescaled mass $\widetilde{M} \equiv M\bra{1+\epsilon}$
and modified horizon coordinate $r_\text{hor}=2M(1+\epsilon)$.
The nonstandard photons perceive a horizon outside the standard
Schwarz\-schild event horizon at $r=r_\text{Schw}\equiv 2M$.\footnote{The
effective background \eqref{eq:SS-lineelementlemaitremodmass} agrees with the
effective metric obtained in Ref.~\cite{DubovskySibiryakov2006} for a
minimally coupled scalar field interacting with the ghost condensate
\cite{ArkaniHamed-etal2004,Mukohyama2005,Feldstein2009}. In the present
article, the background field \eqref{eq:xiexplicit} is introduced
by hand. But it is also possible, as shown in \cite{Betschart-etal2009},
to obtain this background field $\xi^\mu$ by spontaneous symmetry breaking from the
ghost-condensate. For our purpose,  though, it is more convenient to
consider the background \eqref{eq:xiexplicit} as coming
from \emph{explicit} Lorentz violation, avoiding discussion of the stability
of the solution and the related flow of energy or entropy.}
The space lying between these horizons, $2M<r<2M(1+\epsilon)$, will be
referred to as the photonic ergoregion or ergoregion, for short.

\setcounter{equation}{0}
\section{Compton scattering}
\label{sec:Compton}
\subsection{Generalities}
\label{subsec:comptongeneralities}

In this section, we present a concrete realization of the
process proposed by Eling \emph{et al.}~\cite{Eling-etal2007},
which, in an appropriate Lorentz-violating theory,
corresponds to a type of Penrose-mechanism~\cite{Penrose1969,Piran-etal1975}
to extract energy from the photonic ergosphere of a nonrotating
Schwarzschild black hole.

In fact, we consider a
Compton scattering event~\cite{Compton1923,Heitler1954,JauchRohrlich1976,Peskin}
from modified Maxwell theory as defined in Sec.~\ref{sec:Setup}.
Specifically, the theory is given by
the total action \eqref{eq:fullaction} in terms of the Lagrange densities
\eqref{eq:LagMMgrav}, \eqref{eq:LagEH}, and \eqref{eq:LagD},
with Lorentz-violating parameters given by
\eqref{eq:ansatznonbire}, \eqref{eq:ansatzxi}, and \eqref{eq:xiexplicit}.

The scattering event is assumed to take place at
%
%
\begin{eqnarray}\label{eq:scattering-point} 
r_\text{scatter}     &=&2M(1+\epsilon\, \rho)\,,\quad
\theta_\text{scatter} =\pi/2\,,\quad
\phi_\text{scatter}   =0\,,
\end{eqnarray}
with the Schwarzschild mass $M$ from the metric \eqref{eq:SS-lineelementstandard},
the effective Lorentz-violating parameter $\epsilon$ defined by \eqref{eq:epsilon},
and a free parameter $\rho$ taking values between $0$ and $1$.
Using Lema\^{i}tre coordinates \eqref{eq:SS-lineelementlemaitre},
the transformation to a local inertial frame is given by
\begin{eqnarray}
\left({e_{\tau}^{\;0}}\right)_\text{\,scatter}&=&1\,,\quad
\left({e_R^{\;1}}\right)_\text{\,scatter}     =1/\sqrt{1+\epsilon\, \rho}\,,
\nonumber\\
\left({e_\theta^{\;2}}\right)_\text{\,scatter}&=&
\left({e_{\phi}^{\;3}}\right)_\text{\,scatter}= 2 M(1+\epsilon\, \rho)\,,
\label{eq:vierbein_LICS}
\end{eqnarray}
with all other components vanishing.
The asymptotically time-like  
Killing field in local coordinates at
the scattering point \eqref{eq:scattering-point} reads
\begin{equation}\label{eq:Killing-local-coord}
\sigma^a_\text{\,scatter} \equiv e_\mu^{\; a}\,\sigma^\mu\;\Big|_\text{\,scatter}
= \left(1,\, \frac{1}{\sqrt{1+\epsilon\, \rho}},\, 0,\, 0\right)\,.
\end{equation}

\begin{figure}[t]
\begin{center}
\epsfig{file=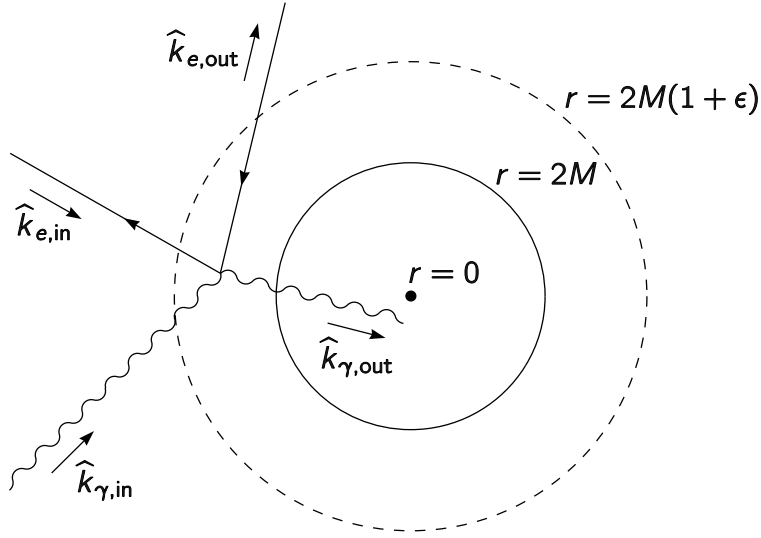,width=0.6\textwidth}
 \caption{Sketch of a Compton scattering event $\gamma e^{-} \to \gamma e^{-}$
 in the photonic ergoregion of a Schwarzschild black hole of mass $M$
 for modified QED \eqref{eq:LagMMgrav}--\eqref{eq:fullaction-LagEH-LagD},
 with Lorentz-violating parameter $\epsilon>0$ defined by \eqref{eq:epsilon}.
 Shown are the unit three-momenta $\widehat{k}_{n}$ of the particles
 ($n=1,\dots,4$) and the flow of positive charge on the electron line.}
 \label{fig:Comptonscat}
\end{center}
\end{figure}

As explained in the Introduction, we are interested in a Compton
scattering event (Fig.~\ref{fig:Comptonscat}) where the
final scattered photon carries negative Killing energy:
\begin{equation}\label{eq:negativeKillingE}
E_{\text{Killing},\gamma,\text{out}}=\sigma^\mu\,
k^\nu_{\gamma,\text{out}}\; \widetilde{g}_{\mu\nu}\equiv \sigma^\mu\,
k_\mu^{\gamma,\text{out}}< 0\,,
\end{equation}
with $k^\nu_{\gamma,\text{out}}$ the tangent vector to the path of
the final photon.
[Here, and in the following, the label `in' or `out' on a particle momentum
refers only to the scattering point and the label `out,'
in particular, does not foretell  the ultimate destiny of the particle.]
Such processes are allowed, since the asymptotically time-like Killing
field for the photon becomes space-like for $r< 2M(1+\epsilon)$.
The final electron should, however, be able to leave to infinity,
carrying more Killing energy than the sum of the initial Killing
energies. [The physical interpretation is that energy is extracted from
the black hole. Thus, it is clear that the complete process is not just an
isolated 2--2 scattering, but that the black hole itself should be considered
as a participant, making this essentially a 3--3
scattering process. However, the treatment as a 2--2 scattering
process in a fixed spacetime background is justified for a
black-hole mass $M$ very much larger than all Killing energies involved.]
Moreover, we demand  that such a \emph{Gedankenexperiment} can be prepared in the
asymptotically flat region of spacetime, i.e., that the two initial
particles come in from spatial infinity.

These conditions impose several constraints on the initial and final
four-vectors of the particles.
For the sake of brevity, these constraints are omitted, but
it has been checked that the example of Sec.~\ref{subsec:example}
fulfills all requirements.

\subsection{Parametrization}\label{subsec:Parametrization}

For our purpose, a useful parametrization of the Compton-scattering wave vectors
(in the local inertial frame with Cartesian coordinates) is given by
\begin{subequations}\label{eq:Ansatz}
\begin{eqnarray}
\big(k_a\big)^{\gamma,\text{out}} &=& E_{\gamma,\text{out}}\,
  \left(1,\, -\zeta \omega_1,\, 0,\, \zeta\, \sqrt{1-\omega_1^2}\right), \\
\big(k_a\big)^{e,\text{out}} &=&
{p}_{e,\text{out}}\,\left(\sqrt{m^2/({p}_{e,\text{out}})^2+1},\,\;\widehat{p}_{e,\text{out}}\right)\,,\\
\big(k_a\big)^{\gamma,\text{in}} &=&
\widetilde{E}_{\gamma,\text{in}}\,
\left(1,\, -\zeta \beta_1,\, -\zeta \beta_2 ,\, s_1\zeta\, \sqrt{1-\beta_1^2-\beta_2^2} \right)\,,\\
k_a^{e,\text{in}} &=& k_a^{e,\text{out}}+k_a^{\gamma,\text{out}}
-k_a^{\gamma,\text{in}}\;,
\end{eqnarray}
\end{subequations}
with arbitrary photon energy $E_{\gamma,\text{out}}>0$,
electron three-momentum $\vec{p}\equiv\left(p_1,p_2,p_3\right)$
$\equiv$ $p_{e,\text{out}}\;\widehat{p}_{e,\text{out}}$
for modulus $p_{e,\text{out}}\equiv |\vec{p}_{e,\text{out}}| >0$,
Lorentz-violating parameter $\zeta\equiv\sqrt{1+\epsilon}> 1$,
and energy $\widetilde{E}_{\gamma,\text{in}}>0$ to be determined
from the dispersion relation of the incoming electron.
The parameters $\omega_1$,  and $\beta_{1,2}$ vary
between $-1$ and $1$, with the additional constraint
$\beta_1^2+\beta_2^2\leq 1$.
The parameter  $s_{1}$ takes the value $+1$ or $-1$.

The \emph{Ansatz} \eqref{eq:Ansatz} ensures that the dispersion
relations for massless Lorentz-violating photons and massive electrons
are fulfilled.

\subsection{Concrete example}\label{subsec:example}

Since the experimental bounds on isotropic Lorentz violation are
tight~\cite{KlinkhamerSchreck2008}, very small
Lorentz violation ($0<\kappa\ll 1$) would be physically more interesting
than large Lorentz violation ($\kappa \sim 1$). However, the Compton scattering
process with negative Killing energy of the final photon
appears to be kinematically forbidden for a small Lorentz-violating parameter
$\kappa$ (see Sec.~\ref{subsec:Small-LV}).

The following example of allowed kinematics is, therefore,
of purely theoretical interest. Specifically, the parameters
are chosen to be
\begin{subequations}\label{eq:example}
\begin{align}
\epsilon&=1/2\,,\quad
&\rho&=99/100\,,
\\
E_{\gamma,\text{out}} &=5\,m\,,\quad
&\omega_1&=9984/10000\,,
\\
p_{e,\text{out}}&=20\, E_{\gamma,\text{out}}\,, \quad
&\widehat{p}_{e,\text{out}}&=
\big(-41,\, 0,\, 3 \sqrt{91} \; \big)\big/50\,,
\\
\beta_1&=74/100,\;\;\;
&\beta_2&=0,\quad s_{1}=1\,,
\end{align}
\end{subequations}
with corresponding  Lorentz-violating parameter
$\kappa=\epsilon/\big(1+\epsilon/2\big)  = 2/5$.
It has taken considerable effort to find this single example.
Apparently, the allowed domain of the multi-dimensional parameter space
is very small, which is confirmed by preliminary numerical calculations.

The above parameters allow for a Compton scattering event that ultimately
reduces the black-hole mass, because the Killing energy of the final photon
is negative: $\sigma^a\,k_a^{\gamma,\text{out}} <0$  using
\eqref{eq:Killing-local-coord} and the above numbers
[the actual value of this energy will be given in
Sec.~\ref{subsec:Gedankenexperiment}].

\subsection{Squared matrix element}

To ensure that the Compton scattering event discussed above has a
nonvanishing probability to occur, the corresponding matrix
element must be nonzero. The squared matrix element for the
Compton scattering process at tree level (calculated with
flat-spacetime electron propagators) reads
\begin{eqnarray}\label{eq:matrix-element}
\hspace*{-10mm}&&
\frac{1}{4}\;\;\sum_{s_1,s_2=\pm 1/2}\;\;\sum_{\lambda_1,\lambda_2=\pm  1}\;\;
|\mathcal{M}|^2=\notag\\
\hspace*{-10mm}&&
\Pi_{ac}\,\Pi_{bd}\;
  \frac{e^4}{4}\;\mathrm{tr}\left\{(\cancel{k}_{e,\text{out}}+m)
  \left[\frac{\gamma^{a}\cancel{k}_{\gamma,\text{in}}\gamma^{b}
  +2\gamma^{a}k_{e,\text{in}}^{b}}{2k_{e,\text{in}}\cdot k_{\gamma,\text{in}}
  +k_{\gamma,\text{in}}^2}
  +\frac{\gamma^{b}\cancel{k}_{\gamma,\text{out}}\gamma^{a}-2\gamma^{b}k_{e,\text{in}}^{a}}
  {2k_{e,\text{in}}\cdot k_{\gamma,\text{out}}
  -k_{\gamma,\text{out}}^2}\right]  \right.
  \notag \\
\hspace*{-10mm}&&
\left. \times  (\cancel{k}_{e,\text{in}}+m)
\left[\frac{\gamma^{d}\cancel{k}_{\gamma,\text{in}}\gamma^{c}
+2\gamma^{c}k_{e,\text{in}}^{d}}{2k_{e,\text{in}} \cdot k_{\gamma,\text{in}}
+k_{\gamma,\text{in}}^2}+\frac{\gamma^{c} \cancel{k}_{\gamma,\text{out}}\gamma^{d}-2\gamma^{d}k_{e,\text{in}}^{c}}
{2k_{e,\text{in}}\cdot k_{\gamma,\text{out}}
-k_{\gamma,\text{out}}^2}\right]\right\}\,,
\end{eqnarray}
with Feynman slash $\cancel{k} \equiv k_a\,\gamma^a$
and photon polarization sum
\begin{equation}
\Pi_{ab}\equiv \sum_{\lambda=\pm 1}
\overline{(\varepsilon^{(\lambda)})}_{a}\,(\varepsilon^{(\lambda)})_{b}\,.
\end{equation}
The Ward identities ensure that,
in gauge-invariant expressions like the one leading up to \eqref{eq:matrix-element},
the polarization sum can be replaced by the following expression
\begin{equation}
\Pi_{ab}  \mapsto
\frac{1}{1+\kappa/2}\left(-\eta_{ab}+\frac{\kappa}{1+\kappa/2}\,\xi_{a}\xi_{b}\right)\,.
\end{equation}
For $k_{\gamma,\text{in}}^2=k_{\gamma,\text{out}}^2=0$ and standard
photon polarization sums, \eqref{eq:matrix-element}
reproduces the standard squared matrix element of Compton scattering;
see, for example, Eq.~(5.81) in Ref.~\cite{Peskin}.

It has now been checked by explicit calculation
that the average squared amplitude \eqref{eq:matrix-element}
is nonzero for the large Lorentz-violating parameter 
and kinematics defined by \eqref{eq:example}.
This particular Compton scattering  event has, therefore,
a nonvanishing probability to occur
[it has also been verified that the same holds
for final photon energies $E_{\gamma,\text{out}}\geq m$,
while keeping the other values in \eqref{eq:example} unchanged].

\subsection{Gedankenexperiment}
\label{subsec:Gedankenexperiment}

At this moment, it may be instructive to give the numerical values of the
four-vectors of the Compton scattering event \eqref{eq:Ansatz}--\eqref{eq:example}:
\begin{subequations}\label{eq:example-4vectors}
\begin{eqnarray}
\big(k_a\big)^{e,\text{in\phantom{n}}}
&\approx&
m\, \left(17.0968,\, -8.44173,\, 0,\, -14.833628\right)\,,
\label{eq:example-4vector-e-in}
\\
\big(k_a\big)^{\gamma,\text{in\phantom{n}}}
&\approx&
m\,\left(87.9082,\, -79.6722,\, 0,\, +72.4163\right)\,,
\label{eq:example-4vector-gamma-in}
\\
\big(k_a\big)^{e,\text{out}}
&\approx&
m\,\left(100.005,\, -82.0000,\, 0,\, +57.2364\right)\,,
\label{eq:example-4vector-e-out}
\\
\big(k_a\big)^{\gamma,\text{out}}
&\approx&
m\, \left(5.00000,\, -6.11393,\, 0,\, +0.346272\right)\,,
\label{eq:example-4vector-gamma-out}
\end{eqnarray}
\end{subequations}
where the three-momenta are seen to lie in a plane ($k_2=0$).
The resulting (conserved) Killing energies of the particles are
\begin{subequations}\label{eq:example-Killing-energies}
\begin{eqnarray}
\big(E_{\text{Killing}}\big)^{e,\text{in\phantom{n}}}
&\approx&\phantom{+}
10.19264 \, m\,,
\\\label{eq:example-Killing-e-in}
\big(E_{\text{Killing}}\big)^{\gamma,\text{in\phantom{n}}}
&\approx&\phantom{+}
22.74743\, m\,,
\\\label{eq:example-Killing-gamma-in}
\big(E_{\text{Killing}}\big)^{e,\text{out}}
&\approx&\phantom{+}
32.94041\,m\,,
\label{eq:example-Killing-e-out}
\\
\big(E_{\text{Killing}}\big)^{\gamma,\text{out}}
&\approx&
- \phantom{0} 0.00034\,  m\,,
\label{eq:example-Killing-gamma-out}
\end{eqnarray}
\end{subequations}
where the energy \eqref{eq:example-Killing-e-out} of the escaping electron
is seen to be larger than the total energy of the two incoming particles,
$E_{\text{Killing}}^{\text{in}} \approx 32.94007\,m$.

A possible \emph{Gedankenexperiment} (in the \emph{Gedankenwelt} of this Letter)
consists of three steps. First, prepare electron and photon beams to give
momenta \eqref{eq:example-4vector-e-in}--\eqref{eq:example-4vector-gamma-in}
at the scattering point \eqref{eq:scattering-point}.
Second, count the number of electrons scattered in the
direction corresponding to \eqref{eq:example-4vector-e-out}
and measure their energy.
Third, determine the change of black-hole mass
(for example, by measuring the change in the orbit of a test particle
encircling the black hole).

\subsection{Small Lorentz violation}
\label{subsec:Small-LV}

A straightforward but tedious analysis for the case of vanishing electron
mass, $m=0$, shows that the  above Compton scattering process
is not allowed for small (but finite) Lorentz-violating  parameter $\epsilon$.
Very briefly, the argument consists of two steps.
First, the dispersion relation for the initial
electron can be solved in terms of the energy of the initial photon.
Second, this initial photon energy can be expanded in $\epsilon$.
For any configuration of the parameters discussed in
Sec.~\ref{subsec:Parametrization}, it can be shown that this photon energy
becomes negative or imaginary for sufficiently small $\epsilon$, if the
constraints mentioned
in the last paragraph of Sec.~\ref{subsec:comptongeneralities}
are taken into account.
It is not easy to get the explicit analytic bound,
but a conservative bound can be found and is given by $\epsilon<1/10$.
That is, it can be shown rigorously that the Compton scattering process
with negative Killing energy of the final photon is kinematically forbidden
for $\epsilon<1/10$.

For the case of nonvanishing electron mass, $m>0$,
numerical investigations show that the process is, once more,
kinematically forbidden for small enough $\epsilon$.
A conservative bound is, again, given by $\epsilon<1/10$
(corresponding to $\kappa<2/21$).

The surprising result, then, is that the reduction of the black-hole mass
by the specific Compton scattering
process appears to be separated from the standard situation
of non-decreasing black-hole mass~\cite{Hawking1971}
by a \emph{finite gap} of the Lorentz-violating parameter $\kappa$.
At the moment, it is not clear if this is just an artefact of the
specific process considered (to be overcome by a more complicated
setup) or if it indicates the existence of a mechanism that
protects the Hawking area theorem~\cite{Hawking1971} for the case
of ``small enough Lorentz violation.''
This interesting question deserves further study.

\section{Discussion}
\label{sec:Discussion}

This Letter investigated the kinematics of
Compton scattering in the accessible part of the photonic ergo\-region of a
Schwarzschild black hole for nonbirefringent modified Maxwell
theory \eqref{eq:LagMMgrav}--\eqref{eq:fullaction-LagEH-LagD}.
More specifically, a Compton scattering event (Fig.~\ref{fig:Comptonscat})
was considered, for which the scattered photon carries away
negative Killing energy \eqref{eq:negativeKillingE}
and ultimately reduces the mass of
the black hole. By giving a concrete example, it has been
shown that such an event is kinematically allowed and has
a nonzero matrix element.

This particular type of Compton scattering event has, therefore,
a nonvanishing probability to occur,
at least, for a relatively large Lorentz-violating parameter $\kappa$.
In a \emph{Gedankenexperiment} starting with a large number $N_\text{BH}$ of
Schwarzschild black holes of identical mass $M$ and having
a large number $N_\text{scatt}$ of repeated Compton scattering events on each
of these black holes, it is then possible to
find certain black holes for which
the initial mass $M$ has been reduced by a macroscopic amount.
In this way, the Hawking area theorem~\cite{Hawking1971}
is circumvented by the presence of negative-energy states
outside the Schwarzschild radius, whose existence is
due to the Lorentz violation of the photonic theory
considered.

These area-reducing events are believed~\cite{Eling-etal2007}
to contradict the generalized second law  of
thermodynamics~\cite{Bekenstein1974}, since they may allow for a
construction of a perpetuum mobile of the second kind. The basic
idea is that such events decrease the mass of the black hole and with
it the associated entropy.
If the scattering were classical~\cite{Eling-etal2007},
the outgoing electron would not carry entropy.\footnote{This would precisely
be the difference with the mining technique of Ref.~\cite{UnruhWald1982},
for which the black-hole mass is also reduced but the
outgoing box (with the mined energy) does carry entropy, namely, that
of the trapped ``acceleration radiation.''}
The whole process would, then, globally \emph{decrease} entropy, in
contradiction with the generalized second law~\cite{Bekenstein1974}.

However, the Compton scattering process discussed above \emph{is}
a quantum process. Certainly, a particular type of Compton
scattering event has been shown to have
a nonvanishing probability to decrease the black-hole mass and
reduce the black-hole entropy. But there is also the possibility that
both particle trajectories after the scattering head towards the black hole
and that the black-hole entropy increases.

An analogous classical process with reduced black-hole entropy
would surely be able
to violate the generalized second law, since it would be possible to
conceive of a deterministic experiment that would result in a decrease of
entropy. But the possibility of an entropy-decreasing quantum
process need not imply, by itself, the violation of the generalized second law.
For example, already in a system with two types of molecules, there is a
nonvanishing  probability that a slow-moving (``cold'') molecule
transfers  energy to a fast-moving (``hot'') molecule. In fact, it is only
the application of statistical mechanics to a system with a large number of molecules
that recovers the second law of thermodynamics~\cite{LandauLifshitzStatMech1}.

A quantitative ana\-ly\-sis would be needed
to see whether or not the Compton scattering process of Sec.~\ref{sec:Compton}
would be able to violate the generalized second law.
This would require phase-space integrations
with nontrivial cuts to determine the probabilities for the
interesting Compton scattering
events to occur. Perhaps one might, then, be able to show a violation of the
fluctuation theorem~\cite{Fluctuation1993,Fluctuation2002}, which might, in turn,
imply the breakdown of the generalized second law.

A more speculative idea expands on the \emph{Gedankenexperiment} discussed
in the last paragraph of Sec.~\ref{subsec:Gedankenexperiment}.
Perhaps it is possible to arrange
for a cloud of electrically charged particles and a pulse of light
coming in from infinity
to scatter elastically at point \eqref{eq:scattering-point} with average
momenta \eqref{eq:example-4vector-e-in}--\eqref{eq:example-4vector-gamma-in}
and to have a final cloud and pulse taking off with
average momenta \eqref{eq:example-4vector-e-out}--\eqref{eq:example-4vector-gamma-out}.
If that arrangement were possible (admittedly a big `if'),
the discussion of Sec.~IV--B  
  of Ref.~\cite{Eling-etal2007}
could be taken over literally, with the consequent
violation of the generalized second law (the incoming and outgoing
charged clouds would have the same velocity dispersion and other
characteristics, the scattering being elastic by assumption).

Whether or not a violation of the generalized second law of
thermodynamics occurs in Lorentz-violating theories remains, therefore, an open
question.\footnote{A different mechanism to violate the
generalized second law was suggested in Ref.~\cite{Feldstein2009}.
A quasi-stationary solution was constructed from the ghost condensate to
describe the flow of negative energy into a black hole.
Just as the process described in the present
article,  this flow of negative energy appears to be able to reduce the
black-hole mass. Potential problems with the stability of this solution
have been discussed and seem to be under control. But,
whether or not this reduction of the black-hole mass results in a violation of
the generalized second law remains, in our opinion, an open question, since
the entropy of the effective ghost-condensate fluid
has not been considered and the issue of turning the flow on and off
requires further analysis. Similar and other reservations regarding the
claim of generalized-second-law violation by Ref.~\cite{Feldstein2009}
have been presented in Ref.~\cite{Mukohyama2009}.}
The present Letter tried to find a concrete realization of
the promising idea~\cite{Eling-etal2007} of exploiting a
Penrose-mechanism-type process.
However, as discussed above, we did not succeed in obtaining an
entirely convincing and totally explicit \emph{Gedankenexperiment}
that is able to violate the generalized second law.
Still, the presented Compton scattering events,  
being able to reduce the black-hole mass,
may provide a step towards demonstrating the violation of
the generalized second law in the Lorentz-noninvariant theory considered,
if at all possible.

\section*{Acknowledgments}
It is a pleasure to thank H. Sahlmann and
S.M. Sibiryakov for helpful comments on the manuscript.



\end{document}